\documentclass[prl,twocolumn,showpacs,superscriptaddress,amsfonts,amsmath,
floatfix]{revtex4}
\usepackage{graphicx}
\usepackage{psfrag}
\usepackage{epsfig}
\usepackage{epstopdf} 
\newcommand{\Journal}[4]{#1 {\bf #2}, #3 (#4)}
\newcommand{\PRL}{Phys. Rev. Lett.}
\newcommand{\PRA}{Phys. Rev. A}
\newcommand{\PRB}{Phys. Rev. B}
\newcommand{\PRD}{Phys. Rev. D}

\newcommand{\NJP}{New J. Phys.}
\newcommand{\FouPh}{Found. Phys}

\newcommand{\Nature}{Nature}

\begin{document}
\title {Number squeezing, quantum fluctuations and oscillations in  mesoscopic Bose Josephson junctions}
\author{G. Ferrini}
\author{A. Minguzzi}
\email{anna.minguzzi@grenoble.cnrs.fr}
\author{F.W.J. Hekking}
\affiliation{Universit\'e Joseph Fourier, Laboratoire de Physique et Mod\'elisation des
Mileux Condens\'es, C.N.R.S. B.P. 166, 38042 Grenoble, France}
\date{\today}

\begin{abstract}
Starting from a quantum two-mode Bose-Hubbard Hamiltonian we determine the ground state
properties, momentum distribution and dynamical evolution for a Bose Josephson junction
realized by an ultracold Bose gas in a double-well trap. Varying the well asymmetry we
identify Mott-like regions of parameters where number fluctuations are suppressed and the
interference fringes in the momentum distribution are strongly reduced. We also show how
Schroedinger cat states, realized from an initially phase coherent state by a sudden rise
of the barrier among the two wells, will give rise to a destructive interference in the
time-dependent momentum distribution.
\end{abstract}
\maketitle
Superconductor Josephson junctions are a paradigmatic example of macroscopic quantum
coherence. The underlying physical mechanism is the Josephson effect~\cite{Josephson63}:
two superconductors connected by a weak link have coherent dynamical behavior determined
by the relative macroscopic phase of the superconducting condensates. Josephson junctions
have been used to discuss fundamental concepts in quantum mechanics \cite{Leggett},
perform precision measurements \cite{Tinkham96}, and are now promising candidates to
implement quantum information devices \cite{makhlin2001}. One important feature of
superconducting Josephson junctions is the possibility to precisely control and adjust
the state of the system by varying external parameters, eg a gate-voltage or magnetic
flux.

Bose Josephson junctions have been only recently proposed \cite{Smerzi1997}, and realized
experimentally \cite{expbosejj}, and many issues remain open. In the simplest
configuration a Bose Josephson junction is realized by confining an ultracold Bose gas in
a double-well potential. This configuration can be described using a
two-mode model in which the bosons occupy the lowest level in each well. In the
``classical'' regime of large particle numbers and weak repulsive interactions the gas is
well described by the Gross-Pitaevskii equation which, within the two-mode approximation,
can be recast in the form of generalized Josephson equations for the time evolution of
the relative phase and population imbalance among the two wells \cite{Smerzi1997}. These equations differ
from the original ones used for superconductor Josephson junctions \cite{Tinkham96} by the presence of  a
nonlinear coupling among the phase and population-imbalance variables. This term
originates from boson-boson interactions in the mean-field approximation and gives rise
to a rich dynamical behavior, displaying self trapping and $\pi$ oscillations
\cite{Smerzi1997}.

In this Letter we focus on the mesoscopic ``quantum'' regime beyond the Gross-Pitaevskii
equation, in the limit of strong interactions and/or smaller values of $N$. This gets
within reach of current experiments \cite{expbosequjj}. As interactions are increased
phase fluctuations become more and more important while number fluctuations are
suppressed; the ground state of the system approaches a regime which can be viewed as a
mesoscopic Mott insulator. During the time evolution the phase coherence first degrades
(``phase diffusion'' \cite{phasediff}), but, in a closed quantum system, periodically
revives, as it has been experimentally demonstrated~\cite{Bloch_2002_b}. At intermediate
times between phase collapse and revival, Schroedinger cat states are predicted to form
\cite{Yurke_Stoler_Tara}, but are not easily observable in superconducting Josephson junctions
\cite{Gerry}.

The quantum behavior of superconducting Josephson junctions is usually accounted for by
the standard phase model \cite{Tinkham96,Leggett_Sols}. This model has been
proposed to study the quantum fluctuations in a Bose Josephson junction
\cite{pita_str}, and has been extended for large particle numbers $N$ with subleading $1/N$ corrections~\cite{Anglin_Smerzi}; however it does not account for large population imbalance among the two 
sides of the junction. In this work we overcome this limitation. Using the quantum
two-mode Bose-Hubbard Hamiltonian we investigate the ground state properties of the
junction which we summarize in a  "phase diagram" obtained by studying number
fluctuations at varying well asymmetry and interaction strength. We characterize the
phase-coherence of the junction by calculating the momentum distribution. We then focus
on the quantum dynamical behavior of the junction in a regime where Schroedinger cat
states occur and show how they affect the time-dependent momentum distribution of the
gas.

{\it Model} We start by the two-mode Bose-Hubbard Hamiltonian
\begin{eqnarray}
\label{Ham2mode}
H=E_1^0 \hat a_1^\dagger \hat a_1 +E_2^0 \hat a_2^\dagger \hat a_2 +\frac{U_1}{2}\hat a_1^\dagger \hat a_1^\dagger \hat a_1 \hat a_1 \nonumber \\ + \frac{U_2}{2}\hat a_2^\dagger \hat a_2^\dagger \hat a_2\hat  a_2 - K(\hat a_2^\dagger \hat a_1 + \hat a_1^\dagger \hat a_2)
\end{eqnarray}
where $\hat a_i,\hat a_i^\dagger$ with $i=1,2$ are bosonic field operators satisfying
$[\hat a_i,\hat a_j^\dagger]=\delta_{ij}$, $E_i^0$ are the energies of the two wells,
$U_i>0$ are the boson-boson repulsive interactions and $K$ is the tunnel matrix element,
ie the Rabi oscillation energy in the case of a non-interacting model. The Heisenberg
equations of motion for this model yield
\begin{eqnarray}
i \hbar \partial_t \hat a_1=E_1^0 \hat a_1+U_1 \hat n_1 \hat a_1-K\hat a_2 \nonumber \\
i \hbar \partial_t \hat a_2=E_2^0 \hat a_2+U_2 \hat n_2 \hat a_1-K\hat a_1,
\end{eqnarray}
where $\hat n_i=\hat a_i^\dagger\hat  a_i $. This is the quantum equivalent of the
two-mode model used in the mean-field approximation $\langle \hat a_i\rangle=\sqrt{N_i}
\exp(i \theta_i)$ and $\langle\hat  n_i\rangle=N_i$ to describe Bose-Josephson junctions
\cite{Smerzi1997}. Indeed, by defining $n=(N_1-N_2)/2$ and $\phi=\theta_2-\theta_1$ the
above equations are readily transformed into the Josephson-like  equations
\begin{eqnarray}
\label{eqs_class}
 \hbar \partial_t n=  - 2 K \sqrt{\left(N/2\right)^2-n^2} \sin\phi\nonumber \\
 \hbar \partial_t \phi= \Delta \tilde E + n U_s + K n \cos\phi/\sqrt{\left(N/2\right)^2-n^2}
\end{eqnarray}
where $\Delta \tilde E=[E_1^0+U_1(N-1)/2-E_2^0-U_2(N-1)/2]$ and $U_s=(U_1+U_2)$. To go
beyond the mean field model, we now transform the Hamiltonian (\ref{Ham2mode}) into the
exact quantum phase one by defining first  the operators \cite{Loudon_book} $\hat a_i
=\sqrt{\hat n_i+1}\hat e^{i\theta_i}$,  $\hat a_i^{\dagger}  =\hat
e^{-i\theta_i}\sqrt{\hat n_i+1}$ and then the relative-phase and relative-number
operators $\hat e^{i\phi}=\hat e^{i\theta_2}\hat e^{-i\theta_1}$ and $\hat n=(\hat
n_1-\hat n_2)/2$. Throughout the paper we work at fixed total particle number $\hat
N=\hat n_1+\hat n_2=N$. In the new variables the Hamiltonian reads (up to a constant
term)
\begin{eqnarray}
\label{Ham_qp}
H&=&\frac{1}{2}U_s (\hat n-n_0)^2-K\sqrt{\frac{N}{2}-\hat n+1}\hat e^{i\phi}\sqrt{\frac{N}{2}+\hat n+1}\nonumber \\
 &-&K\sqrt{\frac{N}{2}+\hat n+1}\hat e^{-i\phi}\sqrt{\frac{N}{2}-\hat n+1}
\end{eqnarray}
with $ n_0=- \Delta \tilde E/U_s$, related to the well asymmetry.

{\it Quasi-classical limit} By using the commutation relations $[\sqrt{N/2+\hat n+1},\hat
e^{i\phi} ] = (\sqrt{N/2+\hat n+1}-\sqrt{N/2+\hat n}) \hat e^{i\phi}$, and
$[\sqrt{N/2-\hat n+1} ,\hat e^{-i\phi} ] = (\sqrt{N/2-\hat n+1}-\sqrt{N/2-\hat n})\hat
e^{-i\phi} $, and by expanding the Hamiltonian (\ref{Ham_qp}) for large $N$ we obtain:
 \begin{eqnarray}
\label{Ham_qclass}
H&=&\frac{U_s }{2}(\hat n-n_0)^2-KN \left(1+\frac{1}{N}-\frac{2 \hat n^2+1/2}{N^2}\right)\hat \cos\phi\nonumber \\&-&K  \frac{2i \hat n}{N}\hat \sin \phi,
\end{eqnarray}
where $\hat \cos\phi=(\hat e^{i\phi} +\hat e^{-i\phi})/2 $ and $\hat \sin\phi=-i(\hat
e^{i\phi} -\hat e^{-i\phi})/2$. The leading term in this expansion corresponds to the 
standard phase model used for superconductors, which reads $H_{SJJ}=E_C (\hat
n-n_g )^2/2-E_J\cos\phi $, with $E_C$ the charging energy, $n_g$ the dimensionless gate charge, and $E_J$ the Josephson energy. Hence we have
$E_C\rightarrow U_s$ and $E_J \rightarrow K N$. Also note that the well asymmetry $n_0$ plays the role of $n_g$, commonly used as an external parameter to control the
state of superconducting junctions. The subleading terms in Eq.~(\ref{Ham_qclass}) correspond to a renormalization of the plasma-Josepshon frequency $\sqrt{U_sKN}$, such that  in the limit $U_s\rightarrow 0 $ tends to  the Rabi frequency
$\omega_R=E_J/N=K$ also present  in the mean field solution. Eq.~(\ref{Ham_qclass}) shows
how mesoscopic effects occur in Bose-Josephson junctions, involving a nonlinear coupling between $\hat n$  and $\phi $.

\begin{figure}
\psfrag{x}{$n_0$}
\psfrag{y}{$\gamma$}

\vspace{-1cm}
{\hspace{-0.5cm}\includegraphics[width=6cm,angle=270]{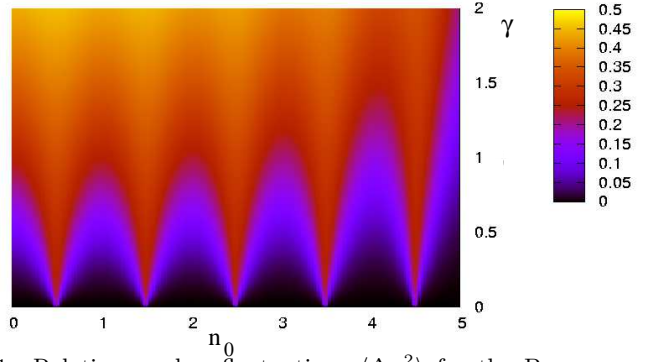}}
\vspace{-1cm}
  \caption{Relative-number fluctuations $\langle \Delta n^2\rangle$ for the
  Bose-Josephson junction with $N=10$ in the plane ($n_0$,$\gamma$).}
  \label{fig1}
\end{figure}

{\it Ground state of the quantum Hamiltonian}.   The quantum Hamiltonian (\ref{Ham_qp})
contains two very different regimes depending on the ratio $\gamma=KN/U_s$: for
$\gamma\gg 1$ it yields a quasi-classical "superfluid" regime, where phase fluctuations
are suppressed and the mean field approximation applies, while for $\gamma\ll 1$ it
yields the fully quantum "Mott-insulator like" regime where number fluctuations are
suppressed (number-squeezed states).

As we work in the two-mode approximation, the quantum Hamiltonian (\ref{Ham_qp}) can be
represented on a finite $(N+1)\times (N+1)$ matrix. In the Fock basis for relative-number
states $|j\rangle $ the Hamiltonian~(\ref{Ham_qp}) has a tridiagonal form with matrix
elements  $\langle j |H| j \rangle=U_s(j-n_0)^2/2$,  $\langle j+1 |H| j \rangle=-K
\sqrt{N/2+j+1}\sqrt{N/2-j}$ and $\langle j-1|H| j \rangle=-K \sqrt{N/2-j+1}\sqrt{N/2+j}$.
We have evaluated the number fluctuations on the ground state of the system numerically,
as represented in Fig.\ref{fig1} in the plane ($n_0,\gamma$). The regions where number
squeezing occurs are reminiscent of the Mott-insulator lobes of the phase diagram of the
Bose-Hubbard model \cite{bose_hubbard}; however we find only a smooth crossover between
the two regimes since we are in the two-mode case. Note that for half-integer values of
$n_0$ the number squeezed regions are strongly suppressed even in the regime $\gamma\ll
1$. At these points the interaction energies of states $j$ and $j+1$ coincide favoring
particle number fluctuations even if $K$ is small. 

While in the limit $\gamma\ll 1$ we have represented the Hamiltonian (\ref{Ham_qp}) in the Fock basis $|j\rangle$ because its eigenvectors are very close to the Fock
states, in the opposite limit $\gamma\gg 1$, where large number fluctuations occur,
 it is more useful to represent the
Hamiltonian by mapping it onto angular momentum variables  in the subspace at fixed
$J^2=N/2(N/2+1)$ \cite{arecchi,walls} (let us choose for simplicity $N$ even). By setting  $\hat
J_x=(\hat a^\dagger_1 \hat a_2+ \hat a^\dagger_2 \hat a_1)/2$, $\hat J_y=-i (\hat
a^\dagger_1 \hat a_2-\hat a^\dagger_2 \hat a_1)/2$, $\hat J_z=(\hat a^\dagger_1 \hat a_1-
\hat a^\dagger_2 \hat a_2)/2=\hat n$ we rewrite the Hamiltonian as:
\begin{equation}
\hat H=U_s (\hat J_z-n_0)^2/2 -2 K \hat J_x.
\end{equation}
Its ground-state eigenvector is close to the angular-momentum coherent state \cite{arecchi}
\begin{equation}
\label{coh_state}
|\alpha\rangle=\sum_{m=-N/2}^{N/2} \left(  \begin{array} {c} N \\ m+N/2 \end{array}\right)^{1/2} \frac{\alpha^{m+N/2}}{(1+|\alpha|^2)^{N/2}} |m\rangle
\end{equation}
with $\alpha=\tan(\theta/2) \exp(-i\phi)$ and $\hat J_z|m\rangle=m|m\rangle$.
Interestingly, the average energy on the state $|\alpha\rangle$ is given by
\begin{equation}
\langle\alpha |\hat H|\alpha\rangle= U_s(1-1/N) n^2/2-2 K \sqrt{(N/2)^2-n^2} \cos\phi
\end{equation}
with $n=-(N/2) \cos\theta$, which corresponds to $O(1/N)$ the mean field result
for the energy, and at $n_0=0$ has a minimal value for $\theta=\pi/2$
and $\phi=0$.

\begin{figure}
  \centering
\psfrag{n(p)}[ ][ ][1][90]{$n(p)$}
\psfrag{pd/(2Pi)}{$pd/2\pi$}
\includegraphics[width=7.5cm,angle=0] {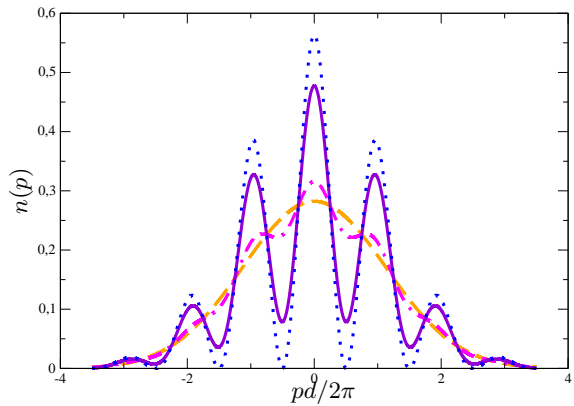}
  \caption{Momentum distribution for a Bose Josephson junction for  various values of the interaction strength ($\gamma=$100,1,0.1 and 0.001 from top to bottom), for $N=10$  and $n_0=0$. For the well wavefunction $\Phi_0(x)$ we have chosen a gaussian profile of width $\sigma=0.1d$.}
  \label{fig2}
\end{figure}

{\it Momentum distribution} The momentum distribution is one of the most accessible
experimental observables. Here we consider the ground-state momentum distribution for the
Bose-Josephson junction using the exact quantum phase model. The field operator in the
two-mode approximation reads $\hat \Psi(x)=\sum_{i=1}^2\Phi_i(x)\hat a_i $, where $\Phi_i(x)$ denotes the ground state wavefunction of the well $i$. The one-body
density matrix, defined as $\rho_1(x,x')=\langle \Psi^\dagger (x) \Psi(x')\rangle$, where
the average is intended  over the quantum state of the system, for the two-mode model
reads $\rho_1(x,x')=\sum_{i,j=1}^2\Phi_j^*(x)\Phi_i(x') \langle a_j^\dagger a_i\rangle $.
The momentum distribution  is the Fourier transform with respect to the relative variable
of the one-body density matrix, $n(p)=\int dx \int dx' \exp(-i p (x-x') )\rho_1(x,x')$
and becomes then $n(p)=\sum_{i,j=1}^2\tilde \Phi_j^*(p)\tilde \Phi_i(p) \langle
a_j^\dagger  a_i\rangle$, with $\tilde \Phi_i(p) $ being the Fourier transform of
$\Phi_i(x) $. For a symmetric well or a weakly asymmetric situation we choose
$\Phi_1(x)=\Phi_0(x-d/2)$, $\Phi_2(x)=\Phi_0(x+d/2)$, with $d$ being the interwell distance,  and hence we obtain for the momentum
distribution
\begin{equation}
\label{mom_distr}
n(p)=|\tilde \Phi_0(p)|^2 (N+e^{-ip d}\langle \hat J_+\rangle+e^{ip d}\langle \hat J_-\rangle).
\end{equation}
This is the generalization of the result derived by  Pitaevskii and Stringari
\cite{pita_str} for a  Bose-Josephson junction using the standard phase model,
and has also been used by Gati et al.~\cite{expbosequjj}  to quantify thermal decoherence in
the experiment. In the quasi-classical regime $\gamma\gg 1$ we can evaluate the average
in Eq.(\ref{mom_distr}) using  the coherent state (\ref{coh_state}), this yields
$n(p)=|\tilde \Phi_0(p)|^2 (N+ 2\sqrt{(N/2)^2-n^2} \cos(pd+\phi)$; here we put $n_0=0$
such that $n=0$ for the ground state. Hence for $\gamma\gg 1$ we expect  interference fringes in momentum
space \cite{pita_str} while in the fully quantum regime $\gamma\rightarrow 0 $ the matrix
elements $\langle \hat J_{+,-}\rangle$ are vanishingly small, and the interferences are
washed out as  illustrated in Fig.\ref{fig2}. Notice that this implies averaging over repeated measurements, as 
a single measurement would still yield interference fringes \cite{CastinDalibard97}.

{\it Schr\"odinger cat states} Schr\"odinger cat states are quantum superpositions of
macroscopic states. We suggest that such states might be realized as a result of the time
evolution following a sudden rise of the barrier between the two wells, starting from an
initially coherent state (ie in the regime $\gamma\gg 1) $. This procedure has the
advantage of starting from a quasi-classical state which is what is currently realized in
experiments. For simplicity, we consider the symmetric case $n_0=0$. If at time $t=0_+$
we set the inter-well coupling $K$ in the Hamiltonian to zero, then the time evolution is
governed by the term $U_s \hat J_z^2/2$ in the Hamiltonian. For each basis vector
$|m\rangle$ of the coherent state, the time evolution is given by $|m(t)\rangle=\exp(-i 2
\pi m^2 t/T)|m\rangle$, where $T=4 \pi \hbar/U_s$ is the revival period \cite{note_times} such that
$|\alpha (T) \rangle = |\alpha \rangle$.  Consider now the special times $T/2 q$, $q$
integer. In this case the phase factor governing the time evolution of the state
$|m\rangle$ becomes $\exp(-i \pi m^2/q )$, which has the (anti-)periodicity property
$\exp(-i \pi (m+q)^2/q ) = (-1)^q \exp(-i \pi m^2/q )$ depending on the parity of $q$.
Hence for even $q$ we can perform a discrete Fourier transform to obtain a cat state
given by a superposition of $q$ coherent states,
\begin{equation}
|\alpha(T/2 q)\rangle = \sum_{k=0}^{q-1} u_k e^{i \pi k N/q} |e^{-i 2 \pi k/q} \alpha
\rangle,
\end{equation}
where $u_k = (1/q) \sum_{m=0}^{q-1} e^{-i \pi m^2/q} e^{i 2 \pi k m/q}$, and
similar cat states exist for odd values of $q$.

\begin{figure}
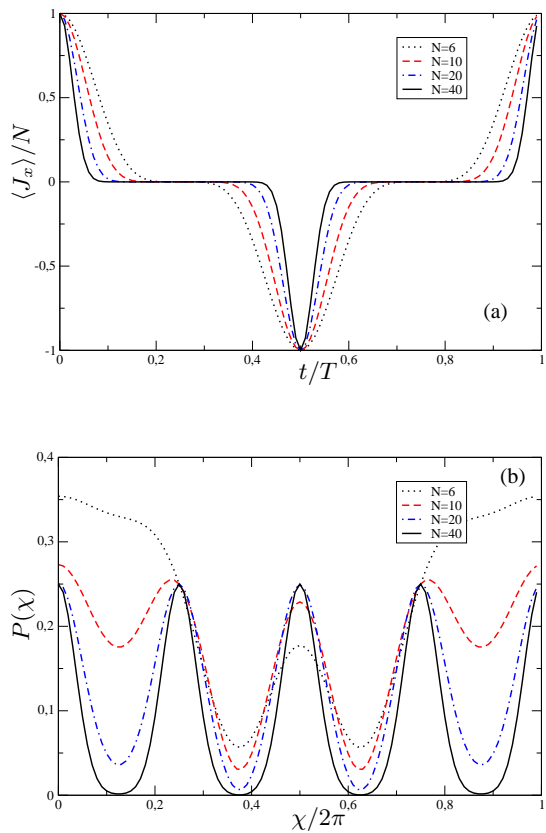

  \centering
\psfrag{t}{$t/T$}
\psfrag{<Jx>/N}{$\langle J_x \rangle/N$}
\includegraphics[width=7cm,angle=0] {fig3a.giulia.eps}
\psfrag{phi}{$\chi/2\pi$}
\psfrag{f(phi)}{$P(\chi)$}

\vspace{0.9cm}{\includegraphics[width=7cm,angle=0]{fig3b.giulia.eps}}
  \caption{Time evolution of $\langle J_x \rangle$  (a), and phase distribution at time $t=T/8$ (b) after a sudden
  quench of the coupling constants starting from an initially coherent state with $\alpha=1$ for various values of $N$.}
  \label{fig3}
\end{figure}

In particular, for $q=2$ we have $|\alpha(T/4)\rangle = [\exp(-i\pi/4)
  |\alpha\rangle+\exp(i\pi/4) (-1)^{N/2} |-\alpha\rangle]/\sqrt{2}$,
which is a cat state given by the superposition of two coherent states. This kind of
state was proposed for light coherent states by Yurke and Stoler \cite{Yurke_Stoler_Tara} and
for Josephson junctions by Gerry \cite{Gerry}. However, in the latter case it was not
easy to probe it. Here, we show that the cat states affect the time-dependent momentum
distribution. In particular, by direct evaluation, we obtain that the contrast in the
momentum distribution vanishes exactly for the two-component cat state. Furthermore, with increasing  $N$, the
contrast gets reduced on larger time regions as higher-order cat states develop at times different from  $t=T/4$  (Fig.\ref{fig3}.a).
For such cat
states the phase distribution $P(\chi)\equiv \langle \alpha (0) \exp(i \chi)|\alpha
(t)\rangle$ has equidistant equal peaks in the interval $[0, 2\pi]$ (Fig.\ref{fig3}.b) which  upon averaging strongly
reduce the momentum-distribution contrast. We have also verified that these features
are robust with respect to small tunneling among the two wells.

{\it Applications and experimental perspectives} Number-squeezed states are particularly
important for atom-optics applications, as their phase-diffusion time is longer than
coherent states (ie usual Bose-condensates) \cite{expbosequjj}. The measurement of the
momentum distribution seems a promising route for the observation of quantum fluctuations
and Schroedinger cat states on Bose Josephson junctions. One difficulty to overcome in
the experimental procedure is the precise control over the number of atoms in the
junction, which should stay constant during the experiment. Atom losses may induce
dephasing as discussed in \cite{decoherence_papers}. On the other side, the time required
for the unitary time evolution into the Schroedinger cat state is experimentally feasible
as demonstrated already with the experiments in optical lattices \cite{Bloch_2002_b}.

\begin{acknowledgments}
We thank C. Bruder, I. Carusotto, L. Glazman and P. Pedri for helpful suggestions. Part
of this work was done at the Institut Henri Poincar\'{e}-Centre Emile Borel  (IHP) in
Paris, during the 2007 workshop ``Quantum Gases''. AM is grateful to the organizers, T.
Leggett, J. Ho, G. Shlyapnikov, and Y. Castin, for the opportunity to participate. We thank IUF 
and CNRS for financial support.
\end{acknowledgments}

\begin{thebibliography}{33}
%
\bibitem{Josephson63} B.D. Josephson, Phys. Lett. {\bf 1}, 251 (1962).
%
\bibitem{Leggett} A.J. Leggett in {\em Chance and matter}, edited by J. Souletie, J. Vannimenus and R.
Stora (Elsevier Science Publishers B.V. 1987).
%
\bibitem{Tinkham96} M. Tinkham, {\em Introduction to Superconductivity} (McGraw-Hill
Singapore, 1996)
%
\bibitem{makhlin2001}Y. Makhlin, G. Sch\"on, and A. Shnirman, Rev. Mod. Phys. {\bf 73},
357 (2001).
%
\bibitem{Smerzi1997} A. Smerzi, S. Fantoni, S. Giovannazzi, and S.R. Shenoy, \Journal{\PRL}{79}{4950}{1997}.
%
\bibitem{expbosejj} M. Albiez  {\it et al.},  \Journal{\PRL}{95}{010402}{2005}; Y. Shin {\it et al.},
\Journal{\PRL}{95}{170402}{2005}; S. Levy {\em et al.},
\Journal{\Nature}{449}{579}{2007}.
%
\bibitem{expbosequjj} R. Gati et al., \Journal{\NJP}{189}{8}{2006}; J. Sebby-Strabley {\it et al.} \Journal{\PRL}{98}{200405}{2007};
G.-B. Jo {\it et al.}, \Journal{\PRL}{98}{030407}{2007}; S. Foelling {\it et al.}, \Journal{\Nature}{448}{1029}{2007}.
%
\bibitem{phasediff} M. Lewenstein and L. You, \Journal{\PRL}{77}{3489}{1996}; Y. Castin and J. Dalibard, \Journal{\PRA}{55}{4330}{1997};
J. Javanein and M. Yu. Ivanov, \Journal{\PRA}{60}{2351}{1999}.
%
\bibitem{Bloch_2002_b} M. Greiner, O. Mandel, T.W. Haensch, I. Bloch,  \Journal{\Nature}{419}{51}{2002}.
%
\bibitem{GorSav99}  D. Gordon and D.M. Savage,  \Journal{\PRA}{59}{4623}{1999}.
%

\bibitem{Yurke_Stoler_Tara} B. Yurke and D. Stoler, \Journal{\PRL}{57}{13}{1986}; D. Stoler \Journal{\PRD}{4}{2309}{1971}, K. Tara, G.S. Agarwal, and S. Chaturvedi, \Journal{\PRA}{47}{5024}{1993}.
%
\bibitem{Gerry} C.C. Gerry \Journal{\PRB}{57}{7474}{1998}.
%
\bibitem{Leggett_Sols} A. Leggett and F. Sols, \Journal{\FouPh}{21}{353}{1991}.
%
\bibitem{bose_hubbard} M.P. Fisher, P.B. Weichman, G. Grinstein, and D.S. Fisher,  \Journal{\PRB}{40}{546}{1989}.
%
\bibitem{pita_str} L. Pitaevskii and S. Stringari, \Journal{\PRL}{87}{180402}{2001}.
%
\bibitem{Anglin_Smerzi} J.R. Anglin, P. Drummond and A. Smerzi,   \Journal{\PRA}{64}{063605}{2001}.
%
\bibitem{Loudon_book} R. Loudon, {\it The Quantum Theory of Light}, (Oxford University, Oxford, 2000).
%
\bibitem{arecchi} F.T. Arecchi, E. Courtens, R. Gilmore, and H. Thomas, \Journal{\PRA}{6}{2211}{1972}.
%
\bibitem{walls} G.J. Milburn, J. Corney, E.M. Wright and D.F. Walls,  \Journal{\PRA}{55}{4318}{1997}.
%
\bibitem{CastinDalibard97} Y. Castin and J. Dalibard \Journal{\PRA}{55}{4330}{1997}.
%
\bibitem{note_times} The two-mode model applies if $T$ is long compared to the inverse of the oscillation frequency of each well.
%
\bibitem{decoherence_papers} P.J.Y. Louis, P.M.R. Brydon, and C.M. Savage, \Journal{\PRA}{64}{053613}{2001} .
\end{thebibliography}
\end{document}